\documentclass{article}

\usepackage{graphicx,amscd,amsmath,amssymb,verbatim}
\usepackage[dvips]{hyperref}
\usepackage{textcomp}
\usepackage[OT1,T1]{fontenc}

\begin{document}

\title{Relativity group for noninertial frames in Hamilton's mechanics}
\author{Stephen G. Low\footnote{Email: Stephen.Low@alumni.utexas.net}}
\date{\today}
\maketitle
\begin{abstract}

The group $\mathcal{E}( 3) =\mathcal{S}\mathcal{O}( 3) \otimes _{s}\mathcal{T}(
3) $, that is the homogeneous subgroup of the Galilei group parameterized
by rotation angles and velocities, defines the continuous group
of transformations between the\ \ frames of inertial particles in
Newtonian mechanics. We show in this paper that the continuous group
of transformations between the frames of noninertial particles following
trajectories that\ \ satisfy Hamilton's equations is given by the
Hamilton group $\mathcal{H}a( 3) =\mathcal{S}\mathcal{O}( 3) \otimes
_{s}\mathcal{H}( 3) $ where $\mathcal{H}( 3) $ is the Weyl-Heisenberg
group that is parameterized by rates of change of position, momentum
and energy, i.e. velocity, force and power.\ \ The group $\mathcal{E}(
3) $ is the inertial special case of the Hamilton group.\ \ 
\end{abstract}
\section{Introduction}

It is very well known that Galilean relativity of inertial frames
with relative velocity and rotation is described by the Euclidean
group that is the homogeneous subgroup of the full Galilei group\footnote{The
full Galelei group also includes translations in position and time
}.\ \ We review, in the following section, the derivation of the
action of the Euclidean group on frames of a particle in Newtonian
space-time from the assumption of invariance of a Newtonian time
line element and invariance of length in the inertial rest frame.
The group multiplication law gives the usual Newtonian addition
of velocity.\ \ The diffeomorphisms of the space-time with these
invariants are the straight lines trajectories of an inertial particle.\ \ 

The group of transformations between frames of particles following
noninertial trajectories also has an invariant Newtonian time line
element and invariance of length in the inertial rest frame.\ \ We
use the Hamilton formulation on extended phase space with position,
time, momentum and energy degrees of freedom and therefore must
also have invariance of the symplectic metric. Using the same method
as reviewed for the Euclidean group, this results in the Hamilton
group that is parameterized by rotation angles, and rates of change
of position, momentum and energy with time, i.e. velocity, force
and power.\ \ The group multiplication law results in the usual
Newtonian addition of velocities and force. The diffeomorphisms
with these invariants must satisfy Hamilton's equations of motion.\ \ The
power transformation law has terms that integrate to those terms
in the Hamiltonian that are required in noninertial frames.\ \ The
homogeneous subgroup of the Galilei group  is the inertial special
case of the Hamilton group. 
\section{Newtonian inertial frames}

 The Newtonian space-time $\mathbb{M}\simeq \mathbb{R}^{n+1}$ has
coordinates $x=(q,t)$ where $q\in \mathbb{R}^{n}$ are the $n$ position
co-ordinates and $t\in \mathbb{R}$ is the time coordinate. The usual
physical case corresponds to $n=3$.\ \ A frame in the cotangent
space at a point $x$, ${{T}^{*}}_{x}\mathbb{M}$, has a basis $d
x=(d q,d t)$. The action of the general linear group element $\Gamma
\in \mathcal{G}\mathcal{L}( n+1,\mathbb{R}) $ on the cotangent space,
suppressing the indices and using basic matrix notation is
\begin{equation}
d \tilde{x }=\Gamma \cdot d x,
\end{equation}

\noindent where $\Gamma $ is a nonsingular\ \ $(n+1)\times (n+1)$
real matrix and $d x$ is a column vector.

The line element may be written as 
\begin{equation}
d s^{2}=d t^{2}={\eta \mbox{}^{\circ}}_{a b}d x^{a}d x^{b} ={}^{t}d
x\cdot \eta \mbox{}^{\circ}\cdot d x,
\end{equation}

\noindent where the indices $a,b..=0,1..n$ and\ \ $\eta  $ is an
$(n+1)\times (n+1)$\ \ matrix\ \ 
\begin{equation}
\eta \mbox{}^{\circ}=\left( \begin{array}{ll}
 0_{n\times n} & 0_{1\times n} \\
 0_{n\times 1} & 1
\end{array}\right) .
\end{equation}

\noindent In this expression, $0_{n\times  m}$ is an $n\times m$
zero matrix.\ \ \ The condition that the line element is invariant
under the action of the group is\ \ 
\begin{equation}
d t^{2}={}^{t}d x\cdot \eta \mbox{}^{\circ}\cdot d x=d {\tilde{t
}}^{2}={}^{t}\left( \Gamma \cdot d x\right) \cdot \eta \mbox{}^{\circ}\cdot
\Gamma \cdot d x ,
\end{equation}

\noindent and therefore
\begin{equation}
\eta \mbox{}^{\circ}={}^{t}\Gamma \cdot \eta \mbox{}^{\circ}\cdot
\Gamma .%
\label{A: Newtonian time line element invariance}
\end{equation}

\noindent We may write $\Gamma $ as an $(n+1)\times (n+1)$ matrix
of the form 
\begin{equation}
\Gamma =\left( \begin{array}{ll}
 R & v \\
 w & \epsilon 
\end{array}\right) 
\end{equation}

\noindent with $R$ an $n\times n$ submatrix, $\epsilon \in \mathbb{R}$
and $v,w\in \mathbb{R}^{n}$ with $v$ a column vector and $w$ a row
vector.\ \ \ Equation (5) results in the expression 
\begin{equation}
\eta \mbox{}^{\circ}=\left( \begin{array}{ll}
 0 & 0 \\
 0 & 1
\end{array}\right) =\left( \begin{array}{ll}
 {}^{t}R & {}^{t}w \\
 {}^{t}v & {}\epsilon 
\end{array}\right)  \left( \begin{array}{ll}
 0 & 0 \\
 0 & 1
\end{array}\right) \left( \begin{array}{ll}
 R & v \\
 w & \epsilon 
\end{array}\right) =\left( \begin{array}{ll}
 {}^{t}w w & {}^{t}w \epsilon  \\
 \epsilon  w & \epsilon ^{2}
\end{array}\right) ,
\end{equation}

\noindent where the dimensions of the zero matrices are now implicit.
It follows directly that $w=0$ and $\epsilon =\pm 1$.\ \ \ 

The group multiplication and inverse property is realized by matrix
multiplication and inverse and a direct calculation shows it defines
the matrix group with group multiplication and inverse given by
\begin{equation}
\begin{array}{l}
 \Gamma ( \epsilon ,R,v) =\Gamma ( \epsilon ^{{\prime\prime}},R^{{\prime\prime}},v^{{\prime\prime}})
\cdot \Gamma ( \epsilon ^{\prime },R^{\prime },v^{\prime }) =\Gamma
( \epsilon ^{{\prime\prime}}\epsilon ^{\prime },R^{{\prime\prime}}\cdot
R^{\prime },R^{{\prime\prime}}\cdot v^{\prime }+\epsilon ^{\prime
}v^{{\prime\prime}}) , \\
 \Gamma ^{-1}( \epsilon ,R,v) =\Gamma ( \epsilon ,R^{-1},-\epsilon
R^{-1}\cdot v) .
\end{array}%
\label{A: Extended inhomogeneos gl multiplication}
\end{equation}
The requirement that $\det  \Gamma \neq 0$ requires that $\det 
R \neq 0$ and therefore $R\in \mathcal{G}\mathcal{L}( n,\mathbb{R})
$.\ \ The group elements $\Gamma ( 1,R,0) $ define the natural embedding
of $\mathcal{G}\mathcal{L}( n,\mathbb{R}) $ into $\mathcal{G}\mathcal{L}(
n+1,\mathbb{R}) $. The elements $\Gamma ( \epsilon ,I_{n},0) \in
\mathcal{D}_{2}$\ \ where $\mathcal{D}_{2}$ is the two element discrete
group of time reversal.\ \ \ Note that $\Gamma ( \epsilon ,R,0)
\in \mathcal{D}_{2}\otimes \mathcal{G}\mathcal{L}( n,\mathbb{R})
$.\ \ \ 

The {\itshape translation} group is defined to be the matrix Lie
group $\mathcal{T}( n) $ that is isomorphic to $\mathbb{R}^{n}$\ \ considered
to be an abelian group under addition, $\mathcal{T}( n) \simeq (\mathbb{R}^{n},+)$.
The group elements $\Gamma ( 1,I_{n},v) $, with {\itshape $I_{n}$}
the $n\times n$ unit matrix,\ \ define elements of the translation
group $\mathcal{T}( n) $ with group composition 
\begin{equation}
\begin{array}{l}
 \Gamma ( 1,I_{n},v) =\Gamma ( 1,I_{n},v^{{\prime\prime}}) \cdot
\Gamma ( 1,I_{n},v^{\prime }) =\Gamma ( 1,I_{n},v^{\prime }+v^{{\prime\prime}})
, \\
 \Gamma ^{-1}( 1,I_{n},v) =\Gamma ( 1,I_{n},-v) .
\end{array}%
\label{A: Velocity translation addition}
\end{equation}

\noindent In this case, the translation group is parameterized by
velocity: the translation are in velocity space rather than position
space. The automorphisms of this translation subgroup are
\begin{equation}
\Gamma ( \epsilon ^{\prime },R^{\prime },v^{\prime }) \cdot \Gamma
( 1,I_{n},v) \cdot \Gamma ^{-1}( \epsilon ^{\prime },R^{\prime },v^{\prime
}) =\Gamma ( 1,I_{n},{\epsilon  }^{\prime }R^{\prime }\cdot v) ,
\end{equation}

\noindent and therefore the translation group is a normal subgroup.
The intersection of this translation subgroup with the subgroup
$\mathcal{D}_{2}\otimes \mathcal{G}\mathcal{L}( n,\mathbb{R}) $
is the identity and the union is the entire group. Therefore, the
group is the extended inhomogeneous general linear group
\begin{equation}
\mathcal{I}\hat{\mathcal{G}}\mathcal{L}( n,\mathbb{R}) \simeq \mathcal{D}_{2}\otimes
_{s}\mathcal{I}\mathcal{G}\mathcal{L}( n,\mathbb{R}) \simeq \mathcal{D}_{2}\otimes
_{s}\mathcal{G}\mathcal{L}( n,\mathbb{R}) \otimes _{s}\mathcal{T}(
n) .
\end{equation}

The above group does not leave length, $d q^{2}$, invariant in the
rest frame.\ \ The rest frame is the special case where $v=0$.\ \ Requiring
that the line element $d q^{2}$ is invariant in the inertial rest
frame
\begin{equation}
d q^{2}={}^{t}d x\cdot \eta ^{q}\cdot d x={}^{t}d x\cdot {}^{t}\Gamma
( \epsilon ,R,0) \cdot \eta ^{q}\cdot \Gamma ( \epsilon ,R,0) \cdot
d x ,
\end{equation}

\noindent results in the condition $\eta ^{q}={}^{t}\Gamma ( \epsilon
,R,0) \cdot \eta ^{q}\cdot \Gamma ( \epsilon ,R,0) $.\ \ This may
be written in matrix notation as
\begin{equation}
\eta ^{q}=\left( \begin{array}{ll}
 I_{n} & 0 \\
 0 & 0
\end{array}\right) =\left( \begin{array}{ll}
 {}^{t}R & 0 \\
 0 & {}\epsilon 
\end{array}\right)  \left( \begin{array}{ll}
 I_{n} & 0 \\
 0 & 0
\end{array}\right) \left( \begin{array}{ll}
 R & 0 \\
 0 & \epsilon 
\end{array}\right) =\left( \begin{array}{ll}
 {}^{t}R \cdot R & 0 \\
 0 & {}0
\end{array}\right)  ,
\end{equation}

\noindent where $I_{n}$ is the $n\times n$ unit matrix.\ \ This
requires that ${}^{t}R =R^{-1}$ and therefore $R\in \mathcal{O}(
n) $.\ \ \ 

Matrices of the form $\Gamma ( \epsilon ,R,v) $ with $\epsilon =\pm
1$, $v\in \mathbb{R}^{n}$ and $R\in \mathcal{O}( n) $\ \ are elements
of the matrix Lie group, generally called the\ \ extended Euclidean
group, $\hat{\mathcal{E}}( n) $ where
\begin{equation}
\hat{\mathcal{E}}( n) \simeq \mathcal{D}_{2}\otimes _{s}\mathcal{O}(
n) \otimes _{s}\mathcal{T}( n) .%
\label{A: ExtendedEuclidean Group}
\end{equation}

\noindent The group multiplication and inverse is given by (8) with
$R\in \mathcal{O}( n) $. 

The orthogonal group may be written as the semidirect product of
the special orthogonal group and a 2 element discrete parity group
${\tilde{\mathcal{D}}}_{2}$ as $\mathcal{O}( n) ={\tilde{\mathcal{D}}}_{2}\otimes
_{s}\mathcal{S}\mathcal{O}( n) $.\ \ Define $\varsigma \in \mathcal{D}_{4}=\mathcal{D}_{2}\otimes
{\tilde{\mathcal{D}}}_{2}$\ \ as the 4 element parity, time reversal
group with elements
\begin{equation}
\varsigma =\left( \begin{array}{ll}
 \tilde{\epsilon }I_{n} & 0 \\
 0 & \epsilon 
\end{array}\right) ,\ \ \epsilon =\pm 1, \tilde{\epsilon }=\pm 1.%
\label{A: PCT n+1 matrix realization}
\end{equation}

\noindent  The extended Euclidean group (14) may be written as
\begin{equation}
\hat{\mathcal{E}}( n) \simeq \mathcal{D}_{2}\otimes _{s}\mathcal{O}(
n) \otimes _{s}\mathcal{T}( n) \simeq \mathcal{D}_{4}\otimes _{s}\mathcal{S}\mathcal{O}(
n) \otimes _{s}\mathcal{T}( n) \simeq \mathcal{D}_{4}\otimes _{s}\mathcal{E}(
n)  .
\end{equation}

\noindent where $\mathcal{E}( n) \simeq \mathcal{S}\mathcal{O}(
n) \otimes _{s}\mathcal{T}(n)$ is the Euclidean matrix Lie group
that is the homogeneous subgroup of the Galilei group.

Consider a transformation $\tilde{x}=\varphi ( x) $ that preserves
the Newtonian time line element and length in the rest frame.\ \ The
matrix of the Jacobian of the transformation is therefore an\ \ element
of the group,\ \ $[\frac{\partial \varphi ( x) }{\partial x}]|_{x}\in
\mathcal{E}( n) .$ The discrete transformations do not need to be
considered as the $\varphi $ are continuous and therefore only the
continuous group is required. Furthermore, we can rotate the coordinates
so that the rotation group need not be considered.\ \ Then, the
Jacobian is an element of the translation normal subgroup of the
Euclidean group, 
\begin{equation}
\left( \frac{\partial \varphi ^{a}( x) }{\partial x^{a}}\right)
=\left( \begin{array}{ll}
 \frac{\partial \varphi ^{i}( t,q) }{\partial q^{j}} & \frac{\partial
\varphi ^{i}( t,q) }{\partial t} \\
 \frac{\partial \varphi ^{0}( t,q) }{\partial q^{j}} & \frac{\partial
\varphi ^{0}( t,q) }{\partial t}
\end{array}\right) =\left( \begin{array}{ll}
 \delta _{i,j} & v^{i} \\
 0 & 1
\end{array}\right) ,
\end{equation}

\noindent where in this expression indices $i,j=1,..n$ are explicit.\ \ With
$v$ constant and ignoring trivial integration constants,\ \ the
transformation equations may be integrated to
\begin{equation}
{\tilde{q}}^{i}=\varphi ^{i}( q,t) =q^{i}+v^{i} t,\ \ \ \ \ \ \ \ \tilde{
t}=\varphi ^{0}( q,t) =t.
\end{equation}

The Euclidean group defines the transformation between inertial
frames in classical Newtonian mechanics. The Euclidean group leaves
invariant $d t$ and therefore in Newtonian physics there is the
notion of absolute time that all observers agree on.\ \ As the frame
is inertial, the rate of change of momentum is zero and the motion
is {\itshape uniform}. Correspondingly, velocity is simply additive
as given by the group laws (8,9).\ \ 
\section{Hamilton group: Newtonian noninertial frames}

The method used above to derive the Euclidean group for inertial
frames may be applied directly to obtain the group of transformations
between general noninertial frames in Hamilton's mechanics. Again,
we require invariance of the Newtonian time line element and also
that length is invariant in the inertial rest frame. As we are using
the Hamilton formulation, we also require invariance of the symplectic
metric.\ \ \ 

For the noninertial case with non zero rate of change of momentum
and position between frames of particle states, consider the space
$\mathbb{P}\simeq \mathbb{R}^{2n+2}$ that has coordinates $z=(p,q,e,t)$.\ \ \ $p,q\in
\mathbb{R}^{n}$ are the $n$ momentum and $n$ position co-ordinates,
$e\in \mathbb{R}$ is the energy coordinate and $t\in \mathbb{R}$
the time coordinate. $n=3$ is the physical case.\ \ A frame at a
point in the cotangent space ${{T}^{*}}_{z}\mathbb{P}$ has a basis
$d z=(d p, d q,d e,d t)$.\ \ \ The action of an element of $\Phi
\in \mathcal{G}\mathcal{L}( 2n+2,\mathbb{R}) $ on the frame is\ \ 
\begin{equation}
d \tilde{z }=\Phi  \cdot d z,
\end{equation}

\noindent where $\Phi $ is a nonsingular\ \ $2(n+1)\times 2(n+1)$
matrix. 

As in the Euclidean case, we consider the subgroup that leaves invariant
the Newtonian time line element $d s^{2}= d t^{2}$.\ \ 
\begin{equation}
d \tilde{z }= d t^{2}={}^{t}d z\cdot \eta \mbox{}^{\circ}\cdot d
z,
\end{equation}

\noindent where $\eta \mbox{}^{\circ}$ is now a $2(n+1)\times 2(n+1)$
singular matrix.\ \ In addition, we again require that the length
$d q^{2}$ be invariant in the inertial rest frame of the particle.\ \ \ 

Hamilton's mechanics has an additional invariant, the symplectic
metric $-d e\wedge d t+\delta _{i,j}d p^{i}\wedge d q^{j}$ where
$i,j..=1,..n$. 

Let $\Phi $ be an $(2n+2) \times  (2n+2)$ nonsingular real matrix
written in terms of submatrices as
\begin{equation}
\Phi =\left( \begin{array}{lll}
 A & b & w \\
 {}^{t}c & a & r \\
 {}^{t}d & f & \epsilon 
\end{array}\right) ,
\end{equation}

\noindent where $A$ is a $2n\times 2 n$ real matrix, $w,b,c,d\in
\mathbb{R}^{n}\text{}$ and $a,r,f,\epsilon \in \mathbb{R}$.\ \ \ From
the analysis in the previous section, we have immediately that the
invariance of the Newtonian time line element requires $\epsilon
=\pm 1$, $d,f=0$ so that $\Phi \in \mathcal{I}\hat{\mathcal{G}}\mathcal{L}(
2n+1,\mathbb{R}) $,\ \ 
\begin{equation}
\Phi =\left( \begin{array}{lll}
 A & b & w \\
 {}^{t}c & a & r \\
 0 & 0 & \epsilon 
\end{array}\right) , \epsilon =\pm 1.%
\label{A: Phi gl form}
\end{equation}

\noindent Furthermore we know that the requirement that the symplectic
metric is invariant requires that $\Phi \in \mathcal{S}p( 2n+2)
$.\ \ The group $\mathcal{G}$ that leaves both the Newtonian line
element and the symplectic group invariant is the intersection of
these two groups
\begin{equation}
\mathcal{G}\simeq \mathcal{S}p( 2n+2) \cap \mathcal{I}\hat{\mathcal{G}}\mathcal{L}(
2n+1) 
\end{equation}

\noindent This group may be explicitly calculated simply by applying
the symplectic condition to the explicit form of the group elements
$\Phi \in \mathcal{I}\hat{\mathcal{G}}\mathcal{L}( 2n+1,\mathbb{R})
$ given in (22).

In the basis $\{d p, d q, d e, d t\}$ the matrix for symplectic
metric ${}^{t}d z\cdot \zeta \cdot d z$\ \ has the form
\begin{equation}
\zeta =\left( \begin{array}{lll}
 \zeta ^{0} & 0 & 0 \\
 0 & 0 & -1 \\
 0 & 1 & 0
\end{array}\right) .
\end{equation}

\noindent where\ \ \ $\zeta \mbox{}^{\circ}$ is the $2n\times 2
n$ matrix\ \ \ 
\begin{equation}
\zeta \mbox{}^{\circ}=\left( \begin{array}{ll}
 0 & I_{n} \\
 -I_{n} & 0
\end{array}\right) ,
\end{equation}

\noindent with $I_{n}$ the $n\times n$ identity matrix.\ \ 

Next, impose the condition that symplectic metric is invariant ${}^{t}\Phi
\cdot \zeta \cdot \Phi =\zeta $ using $\Phi $ defined in (22)
\begin{equation}
\left( \begin{array}{lll}
 {}^{t}A\cdot \zeta \mbox{}^{\circ}\cdot A  & {}^{t}A\cdot  \zeta
\mbox{}^{\circ}\cdot b & {}^{t}A\cdot  \zeta \mbox{}^{\circ}\cdot
w-\epsilon {}c \\
  {}^{t}b\cdot \zeta \mbox{}^{\circ}\cdot A & {}^{t}b\cdot \zeta
\mbox{}^{\circ} \cdot b & {}^{t}b\cdot \zeta \mbox{}^{\circ}\cdot
w-\epsilon  a \\
 {}^{t}w\cdot \zeta \mbox{}^{\circ}\cdot A+\epsilon \ \ {}^{t}c
& {}^{t}w\cdot \zeta \mbox{}^{\circ}\cdot  b+\epsilon  a & {}^{t}w\cdot
\zeta \mbox{}^{\circ}\cdot  w
\end{array}\right) =\left( \begin{array}{lll}
 \zeta \mbox{}^{\circ} & 0 & 0 \\
 0 & 0 & -1 \\
 0 & 1 & 0
\end{array}\right) .
\end{equation}

\noindent First, ${}^{t}A\cdot \zeta \mbox{}^{\circ}\cdot A=\zeta
\mbox{}^{\circ}$ implies that $A\in \mathcal{S}p( 2n) $.\ \ It follows
from ${}^{t}b\cdot \zeta \mbox{}^{\circ}\cdot A=0$ and ${}^{t}A\cdot
\zeta \mbox{}^{\circ}\cdot b=0$ that $b=0$.\ \ Note that ${}^{t}w\cdot
\zeta \mbox{}^{\circ}\cdot w\equiv 0$ as $\zeta \mbox{}^{\circ}$
is antisymmetric.\ \ Then from the terms\ \ ${}^{t}w\cdot \zeta
\mbox{}^{\circ}\cdot b+\epsilon  a=1$ and ${}^{t}b\cdot \zeta \mbox{}^{\circ}\cdot
w-\epsilon  a=-1$,\ \ with $b=0$ we have $a=\epsilon $.\ \ \ Finally,
the remaining equations are\ \ 
\begin{equation}
\ \ \text{}{}c_{ }=\epsilon \ \ \ {}^{t}A\cdot  \zeta \mbox{}^{\circ}\cdot
w ,\ \ \ \ {}\text{}{}^{t}c_{ }= -\epsilon \ \ {}^{t}w\cdot \zeta
\mbox{}^{\circ}\cdot A
\end{equation}

\noindent Noting that ${}^{t}\zeta \mbox{}^{\circ}=-\zeta \mbox{}^{\circ}$,
these two equation are equivalent and therefore the matrix $\Phi
$ takes the form 
\begin{equation}
\Phi ( \epsilon ,A,w,r) =\left( \begin{array}{lll}
 A & 0 & w \\
 -\epsilon \ \ \ {}^{t}w\cdot  \zeta \mbox{}^{\circ}\cdot A & \epsilon
& r \\
 0 & 0 & \epsilon 
\end{array}\right)  
\end{equation}

It follows straightforwardly that this is a matrix group with the
group multiplication realized by matrix multiplication and the group
inverse by matrix inverse 
\begin{equation}
\begin{array}{l}
 \Phi ( \epsilon ,A,w,r) =\Phi ( \epsilon ^{{\prime\prime}},A^{{\prime\prime}},w^{{\prime\prime}},r^{{\prime\prime}})
\cdot \Phi ( \epsilon ^{\prime },A^{\prime },w^{\prime },r^{\prime
}) , \\
 \Phi ^{-1}( \epsilon ,A,w,r) =\Phi ( \epsilon ,A^{-1},-\epsilon
A^{-1}\cdot w,-r) .
\end{array}%
\label{A: HSP group law}
\end{equation}

\noindent where
\begin{equation}
\begin{array}{l}
 \epsilon =\epsilon ^{{\prime\prime}}\epsilon ^{\prime },\ \ \ A=A^{{\prime\prime}}\cdot
A^{\prime }, \\
 w=\epsilon ^{\prime }w^{{\prime\prime}}+A^{{\prime\prime}}\cdot
w^{\prime }, \\
 r=\epsilon ^{{\prime\prime}}r^{\prime }+\epsilon ^{\prime } r^{{\prime\prime}}-\epsilon
^{{\prime\prime}} {}^{t}w^{{\prime\prime}}\cdot \zeta \mbox{}^{\circ}\cdot
A^{{\prime\prime}}\cdot w^{\prime }.
\end{array}%
\label{A: HSP group law components}
\end{equation}

\noindent It is clear that $\Phi ( 1,A,0,0) \in \mathcal{S}p( 2n)
$ and $\Phi ( \epsilon ,I_{2n},0,0) \in \mathcal{D}_{2}$ and further
that $\Phi ( \epsilon ,A,0,0) \in \mathcal{D}_{2}\otimes \mathcal{S}p(
2n) $
\begin{equation}
\Phi ( 1,A,0,0) =\left( \begin{array}{lll}
 A & 0 & 0 \\
 0 & 1 & 0 \\
 0 & 0 & 1
\end{array}\right) ,\ \ \Phi ( \epsilon ,I_{2n},0,0) =\left( \begin{array}{lll}
 I_{2n} & 0 & 0 \\
 0 & \epsilon  & 0 \\
 0 & 0 & \epsilon 
\end{array}\right) .
\end{equation}

The elements of the discrete group $\mathcal{D}_{2}$ change the
sign of the time and energy degrees of freedom together and the
elements of the symplectic group $\mathcal{S}p( 2n) $ are the usual
symplectic transformations on the position and momentum degrees
of freedom.\ \ \ 

Note also that for $A^{{\prime\prime}}=A^{\prime }=I_{2n}$ and $\epsilon
^{{\prime\prime}}=\epsilon ^{\prime }=1$ that the group multiplication
law reduces to
\begin{equation}
\begin{array}{l}
 \begin{array}{rl}
 \Phi ( 1,I_{2n},w,r)  & =\Phi ( 1,I_{2n},w^{{\prime\prime}},r^{{\prime\prime}})
\cdot \Phi ( 1,I_{2n},w^{\prime },r^{\prime }) 
\end{array} \\
 \Phi ^{-1}( 1,I_{2n},w,r) =\Phi ( 1,I_{2n},-w,-r) 
\end{array}%
\label{A: General Heisenberg group operation}
\end{equation}

\noindent where
\begin{equation}
\begin{array}{l}
 w=w^{{\prime\prime}}+w^{\prime }, \\
 r=r^{\prime }+r^{{\prime\prime}}- {}^{t}w^{{\prime\prime}}\cdot
\zeta \mbox{}^{\circ}\cdot w^{\prime }.
\end{array}
\end{equation}

\noindent and therefore\ \ $\Upsilon ( w,r) =\Phi ( 1,I_{2n},w,r)
$ defines a subgroup $\mathcal{H}( n) $ that has the group multiplication
and inverse given by (32) with $w\in \mathbb{R}^{2n}$ and $r\in
\mathbb{R}$. This group is the Weyl-Heisenberg group.

The automorphisms of this subgroup are
\begin{equation}
\begin{array}{rl}
 \Phi ( \epsilon ,A,w,r)  & =\begin{array}{rl}
 \Phi ( \epsilon ^{\prime },A^{\prime },w^{\prime },r^{\prime })
\cdot \Phi ( 1,I_{2n},w^{{\prime\prime}},r^{{\prime\prime}})  &
\cdot \Phi ^{-1}( \epsilon ^{\prime },A^{\prime },w^{\prime },r^{\prime
}) 
\end{array} \\
  & =\Phi ( 1,I_{2n},\epsilon ^{\prime }A^{\prime }\cdot w^{{\prime\prime}},r^{{\prime\prime}}-{}^{t}w^{\prime
}\cdot \zeta \mbox{}^{\circ}\cdot w^{{\prime\prime}}+{}^{t}w^{{\prime\prime}}\cdot
\zeta \mbox{}^{\circ}\cdot w^{\prime }) 
\end{array}%
\label{A: HSp Automorphisms of the Heisenberg Group}
\end{equation}

\noindent Therefore $\mathcal{H}( n) $ is a normal subgroup. The
union of $\text{$\mathcal{H}( n) $}$with $\mathcal{D}_{2}\otimes
\mathcal{S}p( 2n) $ is the full group and the intersection is the
identity. Thus we have the result that the group $\mathcal{G}$ that
leaves the symplectic metric and the Newtonian time line element
invariant is.
\begin{equation}
\mathcal{G}\simeq \mathcal{S}p( 2n+2) \cap \mathcal{I}\hat{\mathcal{G}}\mathcal{L}(
2n+1) \simeq \mathcal{H}\hat{\mathcal{S}}p( 2n) =\mathcal{D}_{2}\otimes
_{s}\mathcal{S}p( 2n) \otimes _{s}\mathcal{H}( n) %
\label{A: HSp definition}
\end{equation}

It is shown in \cite{folland} that this group is the group of linear
automorphisms of the Weyl-Heisenberg group and therefore is the
maximal group with a Weyl-Heisenberg normal subgroup . 
\subsection{Weyl-Heisenberg group}

\noindent The notational change $w=(f,v)$ with $f,v\in \mathbb{R}^{n}$
enables the group operations of the Weyl-Heisenberg group\ \ to
be written in the form
\begin{equation}
\begin{array}{l}
 \begin{array}{rl}
 \Upsilon ( f,v,r)  & =\Upsilon ( f^{{\prime\prime}},v^{{\prime\prime}},r^{{\prime\prime}})
\cdot \Upsilon ( f^{\prime },v^{\prime },r^{\prime })  \\
  & =\Upsilon ( f^{{\prime\prime}}+f^{\prime },v^{{\prime\prime}}+v^{\prime
},r^{{\prime\prime}}+r^{\prime }-f^{{\prime\prime}}\cdot v^{\prime
}+v^{{\prime\prime}}\cdot f^{\prime }) ,
\end{array} \\
 \Upsilon ^{-1}( f,v,r) =\Upsilon ( -f,-v-r) .
\end{array}%
\label{A: Heisenberg group operations}
\end{equation}

\noindent $\Upsilon ( f,v,r) $ may be realized by the matrix group
\cite{Major}
\begin{equation}
\Upsilon ( f,v,r) =\left( \begin{array}{llll}
 I_{n} & 0 & 0 & f \\
 \ \ 0 & I_{n} & 1 & v \\
 v & -f & 1 & r \\
 0 & 0 & 0 & 1
\end{array}\right) ,
\end{equation}

\noindent and it can be directly verified that matrix multiplication
and inverse realizes the group operations given in (36).\ \ 

The Weyl-Heisenberg group itself is the semidirect product of two
translation groups\ \ 
\begin{equation}
\mathcal{H}( n) \simeq \mathcal{T}( n) \otimes _{s}\mathcal{T}(
n+1) .%
\label{A: Heisenberg semidirect product}
\end{equation}

\noindent This may be shown as follows. Consider\ \ the group multiplication
in (36) with $f=r=0$,\ \ 
\begin{equation}
\begin{array}{l}
 \Upsilon ( 0,v^{{\prime\prime}},0) \cdot \Upsilon ( 0,v^{\prime
},0) =\Upsilon ( f,v,r) =\Upsilon ( 0,v^{{\prime\prime}}+v^{\prime
},0) , \\
 \Upsilon ^{-1}( 0,v,0) =\Upsilon ( 0,-v,0) .
\end{array}%
\label{A: Translation subgroup of Heisenberg (V)}
\end{equation}

\noindent Thus, $\Upsilon ( 0,v,0) \in \mathcal{T}( n) $.\ \ As
in the Euclidean case, these translations are parameterized by velocity.
Furthermore, with $v=0$ and $f,r\neq 0$, we have 
\begin{equation}
\begin{array}{l}
 \Upsilon ( f^{{\prime\prime}},0,r^{{\prime\prime}}) \cdot \Upsilon
( f^{\prime },0,r^{\prime }) =\Upsilon ( f,v,r) =\Upsilon ( f^{{\prime\prime}}+f^{\prime
},0,r^{{\prime\prime}}+r^{\prime }) , \\
 \Upsilon ^{-1}( f,0,r) =\Upsilon ( -f,0,-r) .
\end{array}
\end{equation}

\noindent and therefore $\Upsilon ( f,0,r) \in \mathcal{T}( n+1)
$.\ \ These translations are parameterized by force and power. 

 Finally, a special case of the automorphism given in\ \ (34) gives
\begin{equation}
\Upsilon ( f^{\prime },v^{\prime },r^{\prime }) \cdot \Upsilon (
f,0,r) \cdot \Upsilon ^{-1}( f^{\prime },v^{\prime },r^{\prime })
=\Upsilon ( f,0,r-2 f\cdot v^{\prime }) .
\end{equation}

\noindent The translation subgroup $\Upsilon ( f,0,r) \in \mathcal{T}(
n+1) $ of $\mathcal{H}( n) $ is therefore a normal subgroup.\ \ It
may be shown that this is not the case for the translation subgroup
$\Upsilon ( 0,v,0) \in \mathcal{T}( n) $.\ \ Therefore, the Weyl-Heisenberg
group is the semidirect product given in (38).\ \ \ 

The Weyl-Heisenberg group that appears as a subgroup of the group
of transformations between noninertial frames is parameterized by
velocity, force and power. From the\ \ group multiplication given
in (36), velocity and force are simply additive as expected in Newtonian
mechanics. This identification will become clearer in the following
section as well as the meaning of the power transformation law.
\subsection{Hamilton's equations}

We consider now the transformations $\tilde{z}=\varphi ( z) =\varphi
( p,q,e,t) $\ \ that leave the symplectic metric and the Newtonian
line element $d t^{2}$ invariant.\ \ From (35), the continuous group
leaving this invariant is $\mathcal{H}\mathcal{S}p( 2n) $. The Jacobian
of the transformation, $\frac{\partial \varphi ( z) }{\partial z}$,
must be an element of this group.\ \ \ \ Consider first the case
where the Jacobian is an element of the $\mathcal{H}( n) $ subgroup
of $\mathcal{H}\mathcal{S}p( 2n) $. 
\begin{equation}
\left[ \frac{\partial \varphi ^{\alpha }( z) }{\partial z^{\beta
}}\right] |_{z}=\Upsilon ( f,v,r) 
\end{equation}

\noindent Set $z^{\alpha }=\{p^{i},q^{j},e,t\}$ with $\alpha =1,...2n+2$,\ \ $i,j,..=1,..n$.\ \ Then
the above expression can be expanded out to
\begin{equation}
\left( \begin{array}{llll}
 \frac{\partial \varphi ^{i}( z) }{\partial p^{j}} & \frac{\partial
\varphi ^{i}( z) }{\partial q^{j}} & \frac{\partial \varphi ^{i}(
z) }{\partial e} & \frac{\partial \varphi ^{i}( z) }{\partial t}
\\
 \frac{\partial \varphi ^{n+i}( z) }{\partial p^{j}} & \frac{\partial
\varphi ^{n+i}( z) }{\partial q^{j}} & \frac{\partial \varphi ^{n+i}(
z) }{\partial e} & \frac{\partial \varphi ^{n+i}( z) }{\partial
t} \\
 \frac{\partial \varphi ^{2n+1}( z) }{\partial p^{j}} & \frac{\partial
\varphi ^{2n+1}( z) }{\partial q^{j}} & \frac{\partial \varphi ^{2n+1}(
z) }{\partial e} & \frac{\partial \varphi ^{2n+1}( z) }{\partial
t} \\
 \frac{\partial \varphi ^{2n+2}( z) }{\partial p^{j}} & \frac{\partial
\varphi ^{2n+2}( z) }{\partial q^{j}} & \frac{\partial \varphi ^{2n+2}(
z) }{\partial e} & \frac{\partial \varphi ^{2n+2}( z) }{\partial
t}
\end{array}\right) =\left( \begin{array}{llll}
 \delta _{j}^{i} & 0 & 0 & f^{i} \\
  0 & \delta _{j}^{i} & 0 & v^{i} \\
 v^{j} & -f^{j} & 1 & r \\
 0 & 0 & 0 & 1
\end{array}\right) 
\end{equation}

\noindent The solution of these equations requires the $\varphi
^{\alpha }$ to have the form
\begin{equation}
\begin{array}{l}
 {\tilde{p}}^{i}=\varphi ^{i}( p,q,e,t) =p^{i}+{\varphi _{p}}^{i}(
t) , \\
 {\tilde{q}}^{i}=\varphi ^{n+i}( p,q,e,t) =q^{i}+{\varphi _{q}}^{i}(
t) , \\
 \tilde{e}=\varphi ^{2n+1}( p,q,e,t) =e+H( p,q,t) , \\
 \tilde{t}=\varphi ^{2n+2}( p,q,e,t) =t.
\end{array}%
\label{A: Hamilton noinertial transformations}
\end{equation}

\noindent In addition these equations must satisfy
\[
\begin{array}{ll}
 \frac{\partial \varphi ^{n+i}( z) }{\partial t}=v^{i}=\frac{\partial
\varphi ^{2n+1}( z) }{\partial p^{i}}, &  \frac{\partial \varphi
^{2n+1}( z) }{\partial t}=r. \\
 \frac{\partial \varphi ^{i}( z) }{\partial t}=f^{i}=-\frac{\partial
\varphi ^{2n+1}( z) }{\partial q^{i}}, &  
\end{array}
\]

\noindent that on substituting in (44) is\ \ Hamilton's equations\ \ \label{A:
Hamilton equations}
\begin{equation}
\frac{d {\varphi _{q}}^{i}( t) }{d t }=v^{i}=\frac{\partial H( p,q,t)
}{\partial p^{i} }, \frac{d {\varphi _{p}}^{i}( t) }{d t }=f^{i}=-\frac{\partial
H( p,q,t) }{\partial q^{i} },\ \ \frac{\partial H( p,q,t) }{\partial
t }=r.%
\label{A: Hamilton equations}
\end{equation}

From this result, the identification of $v$ with velocity, $f$ with
force and $r$ with power is clear. The group operation describes
the addition of these quantities for transformations between frames
associated with particles following trajectories that satisfy Hamilton's
equations that are generally noninertial. The terms in the power
transformation in the group multiplication law (56,32) integrate
to the terms in the Hamiltonian required for noninertial frames.

On the other hand, if the Jacobian $\frac{\partial \varphi ( z)
}{\partial z}$ is an element of the subgroup $\mathcal{S}p( 2n)
\subset \mathcal{H}\mathcal{S}p( 2n) $, then the transformations
are the usual canonical transformations on momentum-position space.\ \ 
These transformations leave invariant the symplectic metric $\delta
_{i,j}d p^{i}\wedge d q^{j}$ and Hamilton's equations.

Thus, from the condition that the Newtonian time line element $d
t^{2}$ and the condition that\ \ the symplectic metric $\zeta $
are invariant on a $2n+2$ dimensional space, we have derived Hamilton's
equations on $2n$ dimensional phase space and the invariance under
the canonical transformations with Jacobians elements of $\mathcal{S}p(
2n) $.\ \ \ However, viewed on the $2n+2$ dimensional space, the
transformation group is $\mathcal{S}p( 2n) \otimes _{s}\mathcal{H}(
n) $.\ \ This group transforms between the frames associated with
particles following trajectories defined by Hamilton's equations
that are generally noninertial. 
\subsection{The Hamilton group}

Finally, we may also consider\ \ the invariance of the length line
element 
\begin{equation}
d q^{2}=\delta _{i j}d q^{i}d q^{j}={}^{t}d z\cdot \eta ^{q}\cdot
d z,%
\label{A: Length in Hamiltons mechanics}
\end{equation}

\noindent in the inertial rest frame as in the Euclidean case. The
inertial rest frame is defined by $v=f=r=0$ and therefore 
\begin{equation}
{}^{t}\Phi ( 1,A,0,0,0) \cdot \eta ^{q}\cdot \Phi ( 1,A,0,0,0) =\eta
^{q}.
\end{equation}

The $2n \times  2n$ matrix $A\in \mathcal{S}p( 2n) $ may be decomposed
into the four $n \times  n$ submatrices $A_{\mu ,\nu }$ with $\mu
,\nu =1,2$. In the $2n+2$ dimensional space, the $\eta ^{q}$\ \ and
$\Phi ( 1,A,0,0,0) $ are\ \ given by
\begin{equation}
\eta ^{q} =\left( \begin{array}{llll}
 0 & 0 & 0 & 0 \\
 0 & I_{n} & 0 & 0 \\
 0 & 0 & 0 & 0 \\
 0 & 0 & 0 & 0
\end{array}\right) ,\ \ \Phi ( 1,A,0,0,0) =\left( \begin{array}{llll}
 {}A_{1,1} & {}A_{1,2} & 0 & 0 \\
  {}A_{2,1} & {}A_{2,2} & 0 & 0 \\
 0 & 0 & 1 & 0 \\
 0 & 0 & 0 & 1
\end{array}\right) .
\end{equation}

Then the invariance of the length line element (0) in the inertial
rest frame results in 
\begin{equation}
\left( \begin{array}{llll}
 {} {}{}^{t}A_{2,1}\cdot A_{2,1} & {}{}^{t}A_{2,2}\cdot A_{2,1}
& 0 & 0 \\
  {}{}^{t}A_{2,1}\cdot A_{2,2} & {}^{t}A_{2,2}\cdot A_{2,2} & 0
& 0 \\
 0 & 0 & 0 & 0 \\
 0 & 0 & 0 & 0
\end{array}\right) =\left( \begin{array}{llll}
 0 & 0 & 0 & 0 \\
 0 & I_{n} & 0 & 0 \\
 0 & 0 & 0 & 0 \\
 0 & 0 & 0 & 0
\end{array}\right) ,
\end{equation}

\noindent where the dimensions of the zero submatrices are clear
from the context. From this it follows that $A_{2,2}=R\in \mathcal{O}(
n) $ and $A_{2,1}=0$.\ \ 

The matrices $A$ are elements of $\mathcal{S}p( 2n) $ and therefore
${}^{t}A\cdot \zeta \mbox{}^{\circ}\cdot A=\zeta \mbox{}^{\circ}$.
From this it follows that $\ \ A^{-1}=-\zeta \mbox{}^{\circ}\cdot
{}^{t}A\cdot  \zeta \mbox{}^{\circ}$.\ \ Writing the $2n \times
2n$ matrices $A$ in terms of the four $n \times  n$ submatrices
$A_{\mu ,\nu }$ , we have
\begin{equation}
{\left( \begin{array}{ll}
 {}A_{1,1} & {}A_{1,2} \\
  {}A_{2,1} & {}A_{2,2}
\end{array}\right) }^{-1}=-\left( \begin{array}{ll}
 {}0 & {}I_{n} \\
  {}-I_{n} & 0
\end{array}\right) \cdot \left( \begin{array}{ll}
 {}^{t}A_{1,1} & {}^{t}A_{2,1} \\
  {}^{t}A_{1,2} & {}^{t}A_{2,2}
\end{array}\right) \cdot \left( \begin{array}{ll}
 {}0 & {}I_{n} \\
  {}-I_{n} & 0
\end{array}\right) .
\end{equation}

\noindent For the case $A_{2,1}=0$, and $A_{2,2}=R$\ \ the inverse
may be computed and therefore 
\begin{equation}
\left( \begin{array}{ll}
 {}{A_{1,1}}^{-1} & -{}{A_{1,1}}^{-1}\cdot A_{1,2}\cdot R^{-1} \\
  {}0 & {}R^{-1}
\end{array}\right) =\left( \begin{array}{ll}
 {}^{t}R & {}^{t}A_{1,2} \\
  0 & {}^{t}A_{1,1}
\end{array}\right) .
\end{equation}

\noindent Thus\ \ $A_{1,1}={}^{t}R^{-1}$.\ \ Now, as $R\in \mathcal{O}(
n) $, we have that $R^{-1}={}^{t}R$\ \ and so $A_{1,1}={}R$.\ \ The
remaining condition is that
\begin{equation}
-{}{}^{t}R\cdot A_{1,2}\cdot R^{-1}\equiv {}^{t}A_{1,2}\ \ \ \ \ \mathrm{or}\ \ \ \ {}{}^{t}R\cdot
A_{1,2}\equiv - {}^{t}\left( {}^{t}R\cdot A_{1,2}\right) .
\end{equation}

\noindent As this is true for all $R\in \mathcal{O}( n) $,\ \ we
have $A_{1,2}=0$.\ \ This means that $A$ is realized by the $2n
\times  2n$ matrices of the form
\begin{equation}
A=\left( \begin{array}{ll}
 R & 0 \\
 0 & R
\end{array}\right) , R\in \mathcal{O}( n) ,%
\label{A: A reduces to orthogonal}
\end{equation}

\noindent and therefore $A\in \mathcal{O}( n) $. 

This gives the result that the extended Hamilton group $\hat{\mathcal{H}}a(
n) $ is 
\begin{equation}
\hat{\mathcal{H}}a( n) \simeq \mathcal{D}_{2}\otimes _{s}\mathcal{O}(
n) \otimes _{s}\mathcal{H}( n) .
\end{equation}

\noindent An element of the Hamilton group may be written explicitly
in the $(2n +2)\times  (2n+2)$ matrix realization 
\begin{equation}
\Phi ( \epsilon ,R,v,f,r) =\left( \begin{array}{llll}
 R & 0 & 0 & f \\
 0 & R & 0 & v \\
 v & -f & \epsilon  & r \\
 0 & 0 & 0 & \epsilon 
\end{array}\right) .
\end{equation}

\noindent Again, as it is a matrix group, the group multiplication
and inverse is given by matrix multiplication and inverse. Alternatively,
these are just the special case of (0,0) with $A$ given by (0) and
$w=(f,v)$
\begin{equation}
\begin{array}{l}
 \Phi ( \epsilon ,R,f,v,r) =\Phi ( \epsilon ^{{\prime\prime}},R^{{\prime\prime}},f^{{\prime\prime}},v^{{\prime\prime}},r^{{\prime\prime}})
\cdot \Phi ( \epsilon ^{\prime },R^{\prime },f^{\prime },v^{\prime
},r^{\prime }) , \\
 {\Phi ( \epsilon ,R,f,v,r) }^{-1}=\Phi ( \epsilon ,R^{-1},-\epsilon
R\cdot f,-\epsilon  R\cdot v,-\epsilon \ \ r) ,\text{}
\end{array}%
\label{A: Hamilton group composition law}
\end{equation}

\noindent where 
\begin{equation}
\begin{array}{l}
 \epsilon =\epsilon ^{\prime } \epsilon ^{{\prime\prime}},\ \ \ \ \ R=R^{{\prime\prime}}\cdot
R^{\prime },\ \  \\
 f=\epsilon ^{\prime }f^{{\prime\prime}}+R^{{\prime\prime}}\cdot
f^{\prime }, \\
 v=\epsilon ^{\prime }v^{{\prime\prime}}+R^{{\prime\prime}}\cdot
v^{\prime }, \\
 r=\epsilon ^{\prime }r^{{\prime\prime}}+\epsilon ^{{\prime\prime}}(
r^{\prime }-f^{{\prime\prime}}\cdot R^{{\prime\prime}}\cdot v^{\prime
}+v^{{\prime\prime}}\cdot R^{{\prime\prime}}\cdot f^{\prime }) .
\end{array}
\end{equation}

\noindent These are the transformation equations for velocity $v$,
force $f$ and power $r$ under the extended Hamilton group.

Note that for the inertial case with $f=r=0$, that these reduce
to 
\begin{equation}
\begin{array}{l}
 \begin{array}{rl}
 \Phi ( \epsilon ,R,0,v,0)  & =\Phi ( \epsilon ^{{\prime\prime}},R^{{\prime\prime}},0,v^{{\prime\prime}},0)
\cdot \Phi ( \epsilon ^{\prime },R^{\prime },0,v^{\prime },0)  \\
  & =\Phi ( \epsilon ^{\prime } \epsilon ^{{\prime\prime}},0,\epsilon
^{\prime }v^{{\prime\prime}}+R^{{\prime\prime}}\cdot v^{\prime },0)
,
\end{array} \\
 {\Phi ( \epsilon ,R,0,v,0) }^{-1}=\Phi ( \epsilon ,R^{-1},0,-\epsilon
R\cdot v,0) .
\end{array}
\end{equation}

\noindent With the identification $\Gamma ( \epsilon ,R,v) \simeq
\Phi ( \epsilon ,R,0,v,0) $, these are the group multiplication
and inverse laws for the extended Euclidean group given in (0).
Furthermore, noting that for $f=v=0$ that the Weyl-Heisenberg subgroup
reduces to the translation group (0), we have that 
\begin{equation}
\hat{\mathcal{E}}( n) \subset \hat{\mathcal{H}}a( n) .
\end{equation}

\noindent Thus the inertial group, that is the homogenous subgroup
of the Galilei group,\ \ is a special case of the general noninertial
Hamilton group. 

Now, as in the Euclidean case, the orthogonal group can be decomposed
into\ \ the direct product of the two element discrete parity group
and the special orthogonal group,\ \ $\mathcal{O}( n) \simeq {\tilde{\mathcal{D}}}_{2}\otimes
_{s}\mathcal{S}\mathcal{O}( n) $.\ \ The discrete two element parity
group changes the sign of the position and momentum degrees of freedom
together. The final step is to use this decomposition and again
define the 4 element discrete group with elements $\varsigma \in
\mathcal{D}_{4}\simeq {\tilde{\mathcal{D}}}_{2}\otimes \mathcal{D}_{2}$
and restrict $R\in \mathcal{S}\mathcal{O}( n) $.\ \ The $(2n +2)\times
(2n+2)$ matrix realization of the elements $\varsigma \in \mathcal{D}_{4}$\ \ are\ \ \ 
\begin{equation}
\varsigma =\left( \begin{array}{llll}
 \tilde{\epsilon }I_{n} & 0 & 0 & 0 \\
 0 & \tilde{\epsilon }I_{n} & 0 & 0 \\
 0 & 0 & \epsilon  & 0 \\
 0 & 0 & 0 & \epsilon 
\end{array}\right) ,\ \ \epsilon =\pm 1,\tilde{\epsilon }=\pm 1\text{}
\end{equation}

The Hamilton group may then be written\ \ 
\begin{equation}
\hat{\mathcal{H}}a( n) \simeq \mathcal{D}_{4}\otimes _{s}\mathcal{S}\mathcal{O}(
n) \otimes _{s}\mathcal{H}( n) \simeq \mathcal{D}_{4}\otimes _{s}\mathcal{H}a(
n) %
\label{Hamilton inertial velocity equation}
\end{equation}

\noindent where $\mathcal{H}a( n) =\mathcal{S}\mathcal{O}( n) \otimes
_{s}\mathcal{H}(n)$.
\section{Discussion}

We began the discussion in this paper by considering the group leaving
ariance of spacial length in the inertial rest frame resulted in
the extended Euclidean group of transformations that is the homogeneous
subgroup of the Galilei group.\ \ The diffeomorphisms with a Jacobian
that is an element of this group at a given point in the space-time
define the usual linear inertial transformations. The extended Euclidean
group defines the transformations between inertial frames in the
Newtonian formulation. 

We considered next the group that leaves invariant the Newtonian
line element on a time, position, momentum, energy space formulation
that also has a symplectic metric invariant.\ \ If again we require
the invariance of length in the inertial rest frame, the extended
Hamilton group of transformations results.\ \ The diffeomorphisms
with a Jacobian that is an element of this group at a given point
in the space-time define Hamilton's equations. Particles in classical
mechanics follow trajectories that are defined by solutions to Hamilton's
equations. The frames associated with these trajectories are in
general noninertial.\ \ The extended Hamilton group therefore defines
the transformations between general noninertial frames in the Hamilton
formulation. The extended Euclidean transformations are a special
case of the extended Hamilton transformations corresponding to the
inertial case where the rate of change of momentum and energy are
zero.

The Hamilton group multiplication defines the usual addition of
velocity and force. The noninertial transformations of the power
result in terms involving velocity and force appearing in the power
transformation that integrate to the terms required in the Hamiltonian
in a noninertial frame. 

There is nothing fundamentally physical that distinguishes a particle
in an inertial frame as apposed to a noninertial frame. The usual
choice of inertial frames is simply a mathematical expediency to
simplify the analysis. Furthermore, as inertial frames are related
by a group, one expects that noninertial frames in the neighborhood
of the inertial frame to likewise be related by a group. This is
the Hamilton group. In this classical case, the noninertial formulation
does not result in new physical consequences. 

We know that the Euclidean group is the limit of small velocities,
$v/c\rightarrow 0$, of the Lorentz group of special relativity.\ \ The
Lorentz group defines transformations between frames of inertial
particles in special relativity. Clearly, by the above arguments,
there must be a group of transformations for noninertial frames
in the relativistic case.\footnote{This is often assumed to be general
relativity. The equivalence principle results in particles following
geodesics and so all particles in a purely gravitational system
are locally inertial in the curved manifold.\ \ Consider the case
with other forces where gravity is negligible and the problem of
relativistic noninertial frames remains.} This group must have the
Lorentz group as the inertial special case and contract in a well
defined physical limit, that includes small velocities relative
to $c$, to the Hamilton group.\ \ A group that satisfies these properties
and the new physical consequences is discussed in \cite{Low5,Low6}.

\appendix\label{em}\label{Galelei group}\label{grc}

\end{document}